1# On numerical relations playing a role in nuclear fission

G. Mouze and C. Ythier

Faculté des Sciences, Université de Nice, 06108 Nice cedex 2, France

mouze@unice.fr**Abstract**

The key numbers useful for describing the fission process are the mass number $A_{cl}$ of "the primordial cluster" of the fissioning system and the "magic mass numbers" 82 and 126 of the nascent light and heavy fragments. The mean mass number $\overline{A_L}$ and the mean atomic number $\overline{Z_L}$ of the light fragments are linked to the mass number $A_{cl}$ and to the atomic number $Z_{cl}$ of the primordial cluster by simple relationships, $\overline{A_L} = A_{cl} + 68$, $\overline{Z_L} = Z_{cl} + 28$. The value 54 of $\overline{Z_H}$ is predicted by the nucleon phase model.PACS numbers:

25.85.-w: Fission reactions;

25.70.Jj: Fusion and fusion-fission reactions;

21.60 Gx :Cluster models## 1. Introduction. The mass number $A_{cl}$ of the primordial cluster.

James Terrell observed in 1962 that no neutron is emitted from fission fragments of mass 82 and 126, whereas the number of emitted neutrons increases almost linearly above these A-values [1].

G. Mouze and C. Ythier interpreted this important observation in 2008 as revealing that 82 and 126 could be "magic mass numbers". They suggested the existence in the fission process of an intermediary *"nucleon phase"*, in which nucleon shells rather than proton shells and neutron shells are closed at these mass numbers [2-4].

But there exists another mass number playing a role in the fission process, it is the mass number $A_{cl}$ of the "primordial cluster" present in any fissioning system.

This primordial cluster has to be distinguished from the cluster emitted in the phenomenon of cluster-radioactivity.

In fact, there are two kinds of clusters. Those of the first class are "emitted", by a nucleus. It is for example the case of $^{14}C$, emitted by $^{223}Ra$ according to Rose and Jones [5]:

$^{223}Ra \rightarrow \, ^{209}Pb + \, ^{14}C + 31.83$ MeV. 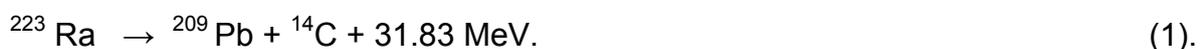 (1).

Their list was recently reviewed by D.N.Poenaru et al. [6]:

$^{14}$C, $^{20}$O, $^{23}$F, $^{22,24-26}$Ne, $^{28,30}$Mg and $^{32,34}$Si. (2)

But the clusters of the second class are not emitted by a nucleus. Their existence was proposed by C. Ythier [7] . He pointed out that if a $^{32}$Mg cluster is formed in the first step of the neutron-induced fission of $^{239}$Pu, according to

$^{239}$Pu + n → $^{208}$Pb + $^{32}$Mg + 79.36 MeV, (3)

the energy released is great enough for inducing a rearrangement process such as:

$^{208}$Pb + $^{32}$Mg → $^{132}$Sn + $^{108}$Ru + 137.7 MeV; (4)

moreover, the clusterization process, for example that described by eq.(3) ,which is strongly energy-yielding , is in fact the reverse of an energy-requiring process, the Oganessian cold fusion reaction, such as the process that led to the discovery of element hassium [8]:

291 MeV—$^{58}$Fe + $^{208}$Pb → $^{265}$Hs + n. (5)

The clusterization process is also comparable to a chemical "equilibrium".

C. Ythier further pointed out that the energy $Q_{cl}$ released by the formation of the "primordial cluster" can be very considerable, e.g. as great as 106.90 MeV for the formation of $^{50}$Ar in the spontaneous fission of $^{258}$Fm, and as great as 241.8 MeV for the formation of $^{78}$Zn in the spontaneous fission of $^{286}$(112). However, this energy $Q_{cl}$ is only equal to 59.49 Mev for the formation of $^{28}$Ne in the neutron-induced fission of $^{235}$U. In fact, a too small value of the clusterization energy can create a new kind of *"fission barrier",* an "energetic" barrier: Such a barrier exists in the case of $^{238}$U, in which the capture of a neutron can induce fission only at a neutron energy greater than 1.5 Mev. Indeed, the clusterization energy released by the formation of a $^{30}$Ne cluster in $^{238}$U is equal to only 45.94 MeV[1]. However, at relativistic energies, $^{238}$U can fission asymmetrically in its collision with a lead target [9],e.g.:

1 Gev -$^{238}$U + Pb → $^{132}$Sn + $^{106}$Mo + 211 MeV, (6)

thanks to the excitation of the $^{238}$U nucleus in the electric field of the 82 protons of the lead nuclei; in such conditions $Q_{cl}$ is equal to about 56.94 MeV:

$^{238}$U $\xrightarrow{+11\ MeV}$ $^{208}$Pb + $^{30}$Ne + 56.94 MeV. (7)

So the non-fissility of $^{238}$U results from an energetic barrier[2], (see, e.g. [10]).

---

[1] 45.941± 0.573 MeV, according to G. Audi et al., Nucl.Phys. A 729,337(2003).
[2] The usual explanation of the non-fissibility of $^{238}$U, given in many textbooks as due to the too small binding energy of the last neutron in $^{239}$U, namely 4.806 MeV, in comparison to the binding energy of the last neutron in $^{236}$U, namely 6.545 MeV, has now to be rejected.

As an example of the importance of $A_{cl}$ and of the magic numbers 82 and 126, we show in Sect.2 how they explain the mass distributions of asymmetric fission, even in regions *far away from symmetry*, as well as the mass distributions of "symmetric" fission.

And as another example of the importance of these numbers, we show in Sect.3 that they are involved in new numerical relationships throwing new light on the nuclear fission process.

## 2. Role of $A_{cl}$ and magic mass numbers in the formation of the mass distributions.

<u>2-1 The asymmetric fission mode</u>

In the nucleon- phase model quoted above, the core-cluster collision made possible by the clusterization energy destroys the $^{208}$Pb- core, which has a much smaller binding energy per nucleon than the cluster. In this destruction, a hard core made of 126 nucleons is assumed to be formed, while <u>82</u> nucleons are released:

$^{208}$Pb  →  A = 126 nucleon-core + <u>82</u> free nucleons         (8)

This hard A = 126 nucleon-core constitutes *the nascent heavy fragment.*

The light fragment originates from the sharing-out of the <u>82</u> free nucleons between primordial cluster and A = 126 nucleon core; but a number, 82 – $A_{cl}$, of them are immediately captured by the cluster, as if it had an enormous tendency to form an A= 82 nucleon core. This A = 82 nucleon core constitutes *the nascent light fragment.*

Light and heavy fragments both result from the partition of the remaining <u>82</u> – (82 – $A_{cl}$) nucleons, i.e. of only $A_{cl}$ nucleons, between the two hard cores [11]. This explains why the regions of appreciate yield extend respectively from A =82 to A = 82 + $A_{cl}$ and from A =126 to A =126 + $A_{cl}$, and why their width is ΔA = $A_{cl}$ for both, as shown in fig.1 [11].

*One sees how essential are here the numbers $A_{cl}$, 82 and 126, as well as the number <u>82</u> of nucleons released by the primordial $^{208}$Pb core !*

<u>2-2 The far-asymmetric mode</u>

 The first region of far asymmetric fission extends from A = $A_{cl}$ to A = 82 (fig.1). This region is formed during the nucleon phase, i.e. within 0.17 yoctosecond, and corresponds to the dressing of the primordial cluster with 82 –$A_{cl}$ nucleons, in order to form the A = 82 hard core.

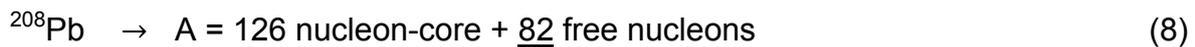

If fragments "smaller than the nascent light fragment" can nevertheless be observed in fission, their yield should be very small, as a consequence of this mode of formation.

This first region has been intensively investigated, in particular at the high flux reactor of the Laue Langevin Institute, Grenoble. It has been found that, between A = 82 and A = 70, the yield, measured at the maximum of the isotopic distribution, drops down over more than five orders of magnitude [13]. However, it must be taken into consideration that the yield essentially depends on the energy of each fragment pair and on its own Coulomb barrier (since in all systems fissioning asymmetrically any fragment pair is confined by its Coulomb barrier [12]), rather than on the mode of formation of the pairs.

The second region of far asymmetric fission, complementary of the first, extends from A = 126 + $A_{cl}$ to A = 208 as shown in fig.1. This region also has been intensively investigated [14].

*One sees again how essential are the numbers $A_{cl}$, 82 and 126,* for the description of the mass distribution of asymmetric fission, even in its far-asymmetric mode.

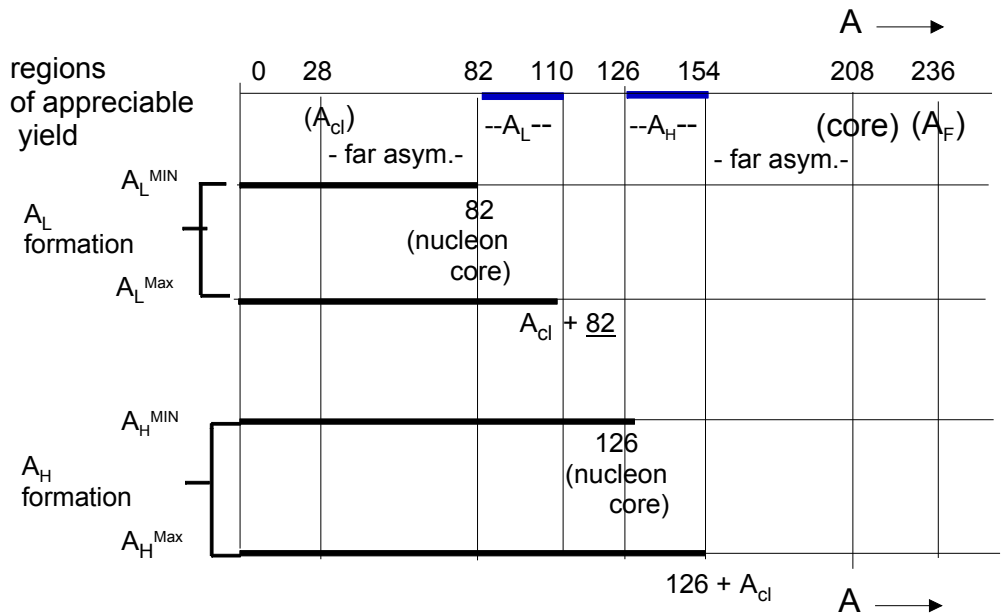

Fig.1 : Explanation of the asymmetric mass distribution of the neutron-induced fission of $^{235}U$.

*It results from the closure of a nucleon shell at A =82 in the light fragment and at A = 126 in the heavy fragment.*

2-3 The symmetric fission mode

The nucleon- phase model explains easily, for the first time, why and how the symmetric fission mode can occur.

Let us consider the system $^{252}$Cf, which fissions asymmetrically. Its primordial cluster is $^{44}$S. If the cluster captures all the 82 nucleons released by the destruction of the $^{208}$Pb core, the heaviest light fragment reaches the mass-value $A_L$ = 126: A hard A =126 nucleon-core can just be formed. It means that *in any system heavier than $^{252}$Cf (s.f.) a light fragment can appear, which is made of a hard A = 126 nucleon core surrounded by one or more valence nucleons*.

Fig.2 shows how the presence of a hard A = 126 nucleon-core in the light fragment allows the creation of a region of appreciable yield of the light fragment which extends from A = 126 to A = $A_{cl}$ + 82. But this region is *precisely the same* as the region of appreciable yield of the heavy fragment, which still extends from A = 126 to A = $A_{cl}$ + 82. Hence the *symmetric* mass distribution. Their common width is also ΔA = $A_{cl}$ – 44. [4].

In $^{286}$(112) (s.f.), the *experimental width* is equal to ΔA + 2, i.e. (78-44) + 2, or 36 mass units[15], as a result of the uncertainty in the mass, due to the extreme brevity of the rearrangement step of the fission process [4,16].

It must be pointed out that in systems fissioning symmetrically a new phenomenon, barrier- free fission, increases considerably the fission yield. Barrier-free fission occurs in systems heavier than $^{258}$Fm (s.f.). In $^{258}$Fm only two fragment pairs, $^{128}$Sn-$^{130}$Sn and $^{126}$Sn-$^{132}$Sn, have an energy of fission greater than their Coulomb barrier. For heavier systems, e.g. $^{286}$(112) (s.f.) [15,16], the yield varies as the differences $E_f$ –$B_c$ calculated for each fragment pair "i" , where $E_f$ is the fission energy and $B_c$ the Coulomb barrier [4,12].

One sees again how essential are the mass-numbers $A_{cl}$ and 126, and the number 82 of nucleons released by the lead primordial core, for the description of the mass distribution of symmetric fission.

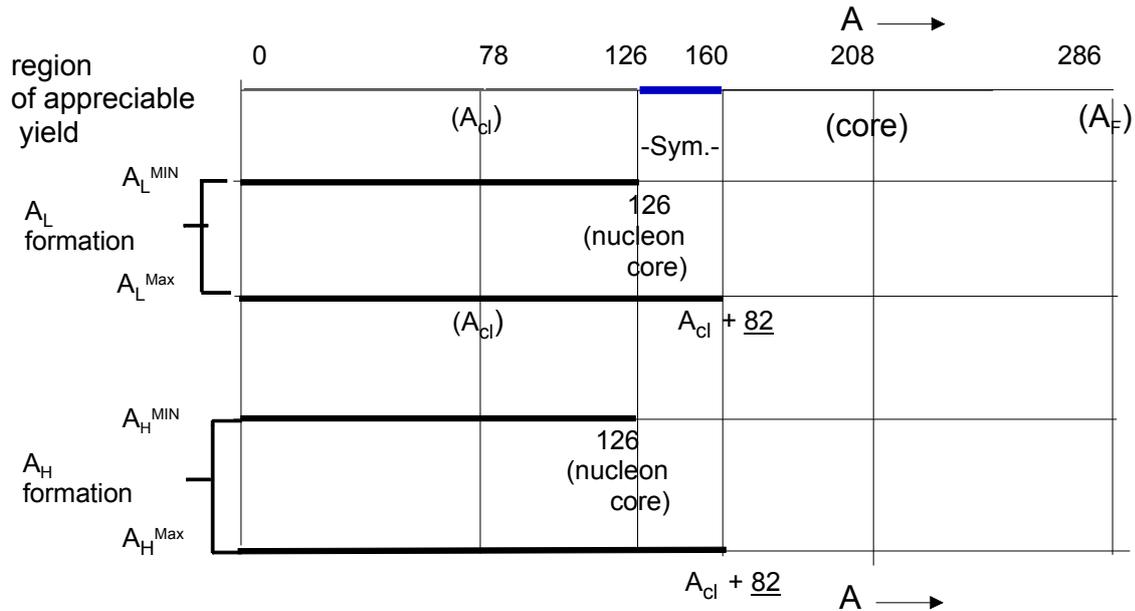

Fig.2: Explanation of the symmetric distribution of the spontaneously fissioning $^{286}$(112) nucleus [15].

*It results from the closure of a nucleon shell at A = 126 in the light fragment.*

### 3. Properties of the mean masses and charges of the fission fragments.

The concept of mean mass number of fission fragments was introduced in the sixties on the occasion of the study of the mass distributions of the fission "products" as a function of the mass of the fissioning systems.

In 1965, A.C. Wahl noted that the mean mass of the heavy fission *products* is almost constant and equal to about 138 [17].

In 1972, K.F. Flynn, L.E. Glendenin and their coworkers [18] reported that the mean mass of the light fission *products* varies linearly as a function of the mass $A_F$ of the the fissioning system.

In 2000, K.-H. Schmidt et al., using the secondary beam facility of G.S.I., Darmstadt, with secondary projectiles produced by fragmentation of the $^{238}$U primary beam, observed that the charge of the heavy "fragment" of all asymmetrically fissioning systems is, on an average, equal to 54 units [19]:

$$\overline{Z_H} = 54. \qquad (9)$$

This important observation of K.-H. Schmidt et al. suggests that the mean value $\overline{Z_L}$ of the proton number $Z_L$ of the complementary *light* fragments of all asymmetrically fissioning systems could be written

$$\overline{Z_L} = Z_F - \overline{Z_H} = Z_F - 54, \quad (10)$$

where $Z_F$ is the proton number of the fissioning system.

In the two-step model developed by G. Mouze and C. Ythier [20], the most energy-rich fragment pair formed in the rearrangement step, e.g. $^{132}$Sn + $^{108}$Ru in eq.(4), can be replaced by a great number of other fragment pairs, and the number of protons transferred to the primordial cluster, instead of being equal to $\Delta Z$ = 32, as in process (4), can have a great number of other values. Consequently, the mean value of the resulting proton number of the light fragments can be written:

$$\overline{Z_L} = Z_{cl} + \overline{\Delta Z} \quad (11)$$

where $\overline{\Delta Z}$ is the mean number of transferred protons.

For the systems $^{233}$U + n, $^{235}$U + n, $^{239}$Pu + n and $^{252}$Cf (s.f.), eq.(10) can be written:

$$\left.\begin{array}{l}\overline{Z_L}\ (^{234}\text{U}) = 92 - 54 = 38 \\[4pt] \overline{Z_L}\ (^{236}\text{U}) = 92 - 54 = 38 \\[4pt] \overline{Z_L}\ (^{240}\text{Pu}) = 94 - 54 = 40 \\[4pt] \overline{Z_L}\ (^{252}\text{Cf}) = 98 - 54 = 44.\end{array}\right\} \quad (12)$$

But eq.(11) can be written:

$$\left.\begin{array}{l}\overline{Z_L}\ (^{234}\text{U}) = 10 + \overline{\Delta Z},\ \text{where 10 is the}\ Z_{cl}\ \text{of the cluster}\ ^{26}\text{Ne} \\[4pt] \overline{Z_L}\ (^{236}\text{U}) = 10 + \overline{\Delta Z},\ \text{where 10 is the}\ Z_{cl}\ \text{of the cluster}\ ^{28}\text{Ne} \\[4pt] \overline{Z_L}\ (^{240}\text{Pu}) = 12 + \overline{\Delta Z},\ \text{where 12 is the}\ Z_{cl}\ \text{of the cluster}\ ^{32}\text{Mg} \\[4pt] \overline{Z_L}\ (^{252}\text{Cf}) = 16 + \overline{\Delta Z},\ \text{where 16 is the}\ Z_{cl}\ \text{of the cluster}\ ^{44}\text{S}.\end{array}\right\} \quad (13)$$

Comparison of eqs.(12) with eqs.(13) leads to the conclusion that

$$\overline{\Delta Z} = 28. \quad (14)$$

*It means that the charge transferred on an average from core to cluster is constant and the same for all fissioning systems.*

We may conclude that to $\overline{Z_H}$ = 54 corresponds a complementary mean value of $Z_L$:

$$\overline{Z_L} = Z_{cl} + 28, \tag{15}$$

according to eq.(11).

For the mean mass of the heavy fragments, we may write:

$$\overline{A_H} \sim 140, \tag{16}$$

since the mean- mass value of the heavy fission *products* has been found equal to ~ 138 for all fissioning systems, and since the mean number of emitted prompt neutrons is, on an average, near to 2 for the same fissioning systems.

In the nucleon-phase model, this $\overline{A_H}$ value can be interpreted as the result of the addition of (140– 126) = 14 nucleons to the nascent heavy fragment made of an A = 126 nucleon-core; and, since 82 nucleons are released in the destruction of the $^{208}$Pb core, the mean number of nucleons transferred to the cluster is necessary equal to 82 – 14 = 68. Hence:

$$\overline{A_H} = 126 + 14 \tag{17}$$

$$\overline{A_L} = A_{cl} + 68 \tag{18}$$

But the last relation is in remarkable agreement with the experimental data of Flynn et al, if these data are represented as a function of $A_{cl}$, instead of $A_F$, and if they are corrected for the corresponding mean number of emitted prompt neutrons, $\bar{\nu}_L$.

If the number of emitted neutrons could be neglected, the mean value $\overline{N_L^{MAX}}$ of the neutron number $N_L$ of the light fragments would be

$$\overline{N_L^{MAX}} = \overline{A_L} - \overline{Z_L} = (A_{cl} + 68) - (Z_{cl} + 28) = N_{cl} + 40. \tag{19}$$

Finally,

$$\left.\begin{array}{l}\overline{A_L} = A_{cl} + 68 \quad \text{and} \quad \overline{A_H} = 126 + 14 = 140 \\ \overline{Z_L} = Z_{cl} + 28 \quad \text{and} \quad \overline{Z_H} = 54 \;\; [19] \\ \overline{N_L^{MAX}} = N_{cl} + 40 \quad \text{and} \quad \overline{N_H^{MAX}} = 86\end{array}\right\} \tag{20}$$

However, if the emission of prompt neutrons is taken into account

$$\overline{N_L} = (N_{cl} + 40) - \bar{\nu}_L \quad \text{and} \quad \overline{N_H} = 86 - \bar{\nu}_H, \tag{21}$$

where $\bar{\nu}_L$ and $\bar{\nu}_H$ are the mean number of the prompt neutrons emitted by the light and heavy fragments.

It is noteworthy that the ratio of the numbers of nucleons transferred to the primordial cluster and to the nascent heavy fragment is

R = 68/14 = 0.206.  (22)

This partition law looks similar to the law established in 1991 by W. Nernst for describing the distribution of a body between two non-miscible phases. The partition law of nuclear fission could also be interpreted as depending on the "chemical potential" of the nucleons in the valence shells of the primordial cluster and of the A = 126 nucleon-core.

## 4. Proof that $\overline{Z_H}$ is equal to 54.

It may be asked whether our hypothesis $\overline{A_H}$ = 54 was necessary, or whether this mean value is in fact a consequence of the nucleon-phase model.

In this model, the formation of hard cores made of 82 and 126 nucleons, i.e. the fact that shell closures appear at these magic "mass numbers", can be explained by the disappearance of the Coulomb field of the protons, which, in ordinary nuclear matter, is responsible for the particular organization of the proton-shells resulting from the spin-orbit coupling.

It means that even if the charge is again present in the fragments at the end of the nucleon phase, *the nucleon-phase itself may be characterized by the absence of any charge*: No change of charge is conceivable during the nucleon transfer. And the ratio Z/A at the end of the transfer is still the same as before the transfer.

But, in the $^{208}$Pb core, the ratio Z/A = 82/208 = 0.39. If Z/A is still the same in the heavy fragment, $(\overline{Z}/\overline{A})_H$ is still equal to 0.39 at the end of the transfer. Thus, since $\overline{A_H}$ = 140, $\overline{Z_H}$ = 0.39 X 140 = 54.

## 5. Discussion

In a recent paper [21] ,C. Böckstiegel et al., assuming that a fissioning system follows "specific valleys in the potential energy in the direction of the elongation", have analyzed the charge distribution of the fission fragments for the following systems: $^{233}$U, $^{232}$Pa, $^{228}$Pa, $^{228}$Th, $^{226}$Th and $^{223}$Th as a function of the "fission channels". This concept of "fission channel" has been introduced by V. Pashkevich [22] and by Brosa et al. [23] in the so- called "independent fission channel model". The most salient observation of Böckstiegel et al. is that the position of the heavy fission fragments of asymmetric fission channels does not vary in atomic number. Thus $\overline{Z_H}$ = ~ 53 for the « standard I »fission channel, whereas $\overline{Z_H}$ = ~ 55 for the "standard II" fission channel.

In our opinion, the different types of "fission channels" are simply epiphenomena resulting from the fact that the various aspects of the mass distributions result from the variation of the fission yield as a function of the difference ( $Q(i) - B_c(i)$ ) calculated for the various fission pairs i, because asymmetric fission is strongly "confined", as indicated above in Sect.2-2.

Thus these various aspects of the mass distributions are not the result of the mode of formation of the fragments. *They reveal only the manner in which $(Q - B_c)$ varies as a function of the fragment mass.*

## 6. Conclusion

We hope to have shown the impressive role played by the mass number $A_{cl}$ and the magic mass numbers 82 and 126 in the description of the mechanism of nuclear fission and of its "various mass distributions"

The observation reported by K.H. Schmidt and coworkers is certainly important, as are also the various new relationships involving $A_{cl}$ related to this $\overline{Z_H}$- value.

But the formation of the light fragment is, on an average, nothing else but the "dressing" of the primordial cluster with 68 nucleons [20]. The transfer of these nucleons occurs within an extremely short time interval [2,3], in conditions characterized by the complete disappearance of the Coulomb field of the protons. It is this very particular situation which justifies the modern expression of Flynn's law and the validity of its new corollaries, as the constancy of $\overline{Z_H}$ and the relationships involving $A_{cl}$.